# Near-Optimal Detection in MIMO Systems using Gibbs Sampling

Morten Hansen*, Babak Hassibi†, Alexandros G. Dimakis††, and Weiyu Xu†

*Technical University of Denmark, Informatics and Mathematical Modelling,
Build. 321, DK-2800 Lyngby, Denmark,
E-mail: mha@imm.dtu.dk

†California Institute of Technology, Department of Electrical Engineering,
Pasadena, CA 91125, USA
†† University of Southern California, Department of Electrical Engineering,

Email: hassibi@systems.caltech.edu, dimakis@usc.edu, and weiyu@caltech.edu

**Abstract**

In this paper we study a Markov Chain Monte Carlo (MCMC) Gibbs sampler for solving the integer least-squares problem. In digital communication the problem is equivalent to performing Maximum Likelihood (ML) detection in Multiple-Input Multiple-Output (MIMO) systems. While the use of MCMC methods for such problems has already been proposed, our method is novel in that we optimize the ”temperature” parameter so that in steady state, i.e. after the Markov chain has mixed, there is only polynomially (rather than exponentially) small probability of encountering the optimal solution. More precisely, we obtain the largest value of the temperature parameter for this to occur, since the higher the temperature, the faster the mixing. This is in contrast to simulated annealing techniques where, rather than being held fixed, the temperature parameter is tended to zero. Simulations suggest that the resulting Gibbs sampler provides a computationally efficient way of achieving approximative ML detection in MIMO systems having a huge number of transmit and receive dimensions. In fact, they further suggest that the Markov chain is rapidly mixing. Thus, it has been observed that even in cases were ML detection using, e.g. sphere decoding becomes infeasible, the Gibbs sampler can still offer a near-optimal solution using much less computations.

## I. Introduction

The problem of performing Maximum Likelihood (ML) decoding in digital communication has gained much attention over the years. One method to obtain the ML solution is Sphere Decoding (SD) [1]–[5]. Over a wide range of Signal-to-Noise Ratios (SNR)s the average complexity of SD is significantly smaller than exhaustive search detectors, but in worst case the complexity is still exponential [6]. Thus, in scenarios with poor SNR or in Multiple-Input Multiple-Output (MIMO) systems with huge transmit and receive dimensions, even SD can be infeasible. A way to overcome this problem is to use approximate Markov Chain Monte Carlo (MCMC) detectors instead, which asymptotically can provide the optimal solution, [7], [8]. Gibbs sampling (also known as Glauber dynamics) is one MCMC method, which is used for sampling from distributions of multiple dimensions. The Gibbs sampler has among others been proposed for detection purposes in wireless communication in [9]–[12] (see also the references therein). The scope of this paper is to describe and analyse a new way of solving the integer least-squares problem using MCMC. It will be shown that the method can be used for achieving a near-optimal and computationally efficient solution of the problem, even for systems having a huge dimension.

The paper is organized as follows; In Section II we present the system model that will be used throughout the paper. The MCMC method is described in Section III and in Section IV we analyse the probability of error for the ML detector. Section V treats the optimal selection of the temperature parameter $\alpha$, while the simulation results are given in Section VI and some concluding remarks are found in Section VII.

## II. System Model

We consider a real-valued block-fading MIMO antenna system, with $N$ transmit and $N$ receive dimensions, with know channel coefficients.[1] The received signal $\mathbf{y} \in \mathbb{R}^N$ can be expressed as

$$\mathbf{y} = \sqrt{\frac{\text{SNR}}{N}}\mathbf{H}\mathbf{s} + \boldsymbol{\upsilon} \ , \tag{1}$$

where $\mathbf{s} \in \Omega^N$ is the transmitted signal, and $\Omega$ denotes the constellation set. To simplify the derivations in the paper we will assume that $\Omega = \{\pm 1\}$. $\boldsymbol{\upsilon} \in \mathbb{R}^N$ is the noise vector where each entry is Gaussian $\mathcal{N}(0,1)$ and independent identically distributed (i.i.d.), and $\mathbf{H} \in \mathbb{R}^{N\times N}$ denotes the channel matrix with i.i.d. $\mathcal{N}(0,1)$ entries. The normalization in (1) guarantees that SNR represents the signal-to-noise ratio per receive dimension (which we define as the ratio of the total transmit energy per channel use divided by the per-component noise variance as described in among others [5]). As explained further below, for analysis purposes we will focus on the regime where $\text{SNR} > 2\ln(N)$, in order to get the probability of error of the ML detector to go to zero. Further, in our analysis, without loss of generality, we will assume that the all minus one vector was transmitted, $\mathbf{s} = -\mathbf{1}$. Therefore

$$\mathbf{y} = \boldsymbol{\upsilon} - \sqrt{\frac{\text{SNR}}{N}}\mathbf{H}\mathbf{1} \ . \tag{2}$$

We are considering a minimization of the average error probability $P(\mathbf{e}) \triangleq P(\hat{\mathbf{s}} \neq \mathbf{s})$, which is obtained by performing Maximum Likelihood Sequence Detection (here simply referred to as ML detection) given by

$$\hat{\mathbf{s}} = \arg\min_{\mathbf{s}\in\Omega^N} \left\| \mathbf{y} - \sqrt{\frac{\text{SNR}}{N}}\mathbf{H}\mathbf{s} \right\|^2 \ . \tag{3}$$

## III. Gibbs Sampling

One way of solving the optimization problem given in (3) is by using Markov Chain Monte Carlo (MCMC) simulations, which asymptotically converge to the optimal solution [13]. More specifically, the MCMC detector we investigate here is the Gibbs sampler, which computes the conditional probability of each symbol in the constellation set at the $j$th index in the estimated symbol vector. This conditional probability is obtained by keeping the $j-1$ other values in the estimated symbol vector fixed. Thus, in $k$th iteration the probability of the $j$th symbol adopts the value $\omega$, is given as

$$p\left(\hat{\mathbf{s}}_j^{(k)}=\omega\,|\theta\right)=\frac{e^{-\frac{1}{2\alpha^2}\left\|\mathbf{y}-\sqrt{\frac{\text{SNR}}{N}}\mathbf{H}\tilde{\mathbf{s}}_{j|\omega}\right\|^2}}{\sum\limits_{\tilde{\mathbf{s}}_{j|\tilde{\omega}}\in\Omega} e^{-\frac{1}{2\alpha^2}\left\|\mathbf{y}-\sqrt{\frac{\text{SNR}}{N}}\mathbf{H}\tilde{\mathbf{s}}_{j|\tilde{\omega}}\right\|^2}}\ , \tag{4}$$

where $\tilde{\mathbf{s}}_{j|\omega}^T \triangleq \left[\hat{\mathbf{s}}_{1:j-1}^{(k)},\omega,\hat{\mathbf{s}}_{j+1:N_T}^{(k-1)}\right]^T$ and where we for simplicity have introduced $\theta=\left\{\hat{\mathbf{s}}^{(k-1)},\mathbf{y},\mathbf{H}\right\}$.[2] $\alpha$ represents a tunable positive parameter which controls the mixing time of the Markov chain, this parameter is also sometimes called the "temperature". The larger $\alpha$ is the faster the mixing time of the Markov chain will be, but as we will show in the paper, there is an upper limit on $\alpha$, in order to ensure that the probability of finding the optimal solution in steady state is not exponentially small. The MCMC method will with probability $p\left(\hat{\mathbf{s}}_j^{(k)}=\omega\,|\theta\right)$ keep $\omega$ at the $j$'th index in estimated symbol vector, and compute conditional probability the $(j+1)$th index in a similar fashion. We define one iteration of the Gibbs sampler as a randomly-ordered update of all the $j=\{1,\ldots,N_T\}$ indices in the estimated symbol vector $\hat{\mathbf{s}}$.[3] The initialization of the symbol vector $\hat{\mathbf{s}}^{(0)}$ can either be chosen randomly or, alternatively, e.g. the zero-forcing solution can be used.

[1]For simplicity we have assumed that the receive and the transmit dimensions are the same, but the results presented in the paper can be generalized to cover different dimensions.

### A. *Complexity of the Gibbs sampler*

The conditional probability for the $j$'th symbol in (4) can be computed efficiently by reusing the result obtained for the $j-1$'th symbol, when we evaluate $\left\|\mathbf{y}-\sqrt{\text{SNR}/N}\mathbf{H}\tilde{\mathbf{s}}_{j|\omega}\right\|^2$. Since we are only changing the $j$'th symbol in the symbol vector, the difference $\mathbf{d}_j \triangleq \mathbf{y}-\sqrt{\text{SNR}/N}\mathbf{H}\tilde{\mathbf{s}}_{j|\omega}$ can be expressed as

$$\mathbf{d}_j=\mathbf{d}_{j-1}-\sqrt{\frac{\text{SNR}}{N}}\mathbf{H}_{1:N,j}\Delta s_{j|\omega}\ , \tag{5}$$

where $\Delta s_{j|\omega}\triangleq s_{j|\omega}^{(k)}-s_{j|\tilde{\omega}}^{(k-1)}$. Thus, the computation of conditional probability of certain symbol in the $j$'th position costs $2N$ operations, where we define an operation as a Multiply and Accumulate (MAC) instruction.[4] This leads to a complexity of $\mathcal{O}\left(2N^2[|\Omega|-1]\right)$ operations per iteration. For further details on the implementation of the Gibbs sampler see [14].

[2]When we compute the probability of symbol $\omega$ at the $j$'th position, we more precisely condition on the symbols $\hat{\mathbf{s}}_{1:j-1}^{(k)}$ and $\hat{\mathbf{s}}_{j+1:N_T}^{(k-1)}$, but to keep the notation simple, we do not explicitly state that in the equations above.

[3]We need a randomly-ordered update for the Markov chain to be reversible and for our subsequent analysis to go through. It is also possible to just randomly select a symbol $j$ to update, without insisting that a full sequence be done. This also makes the Markov chain reversible and has the same steady state distribution. In practice a fixed, say sequential, order can be employed, although the Markov chain is no longer reversible. Note that our theoretical analysis is assuming randomly selected symbol updates for analytical convenience. In our experimental section we used a sequential updating order which empirically yields a slight convergence acceleration.

[4]We need to compute both the inner product $\mathbf{d}_j^T\mathbf{d}_j$ and the product $\mathbf{H}_{1:N,j}\Delta s_{j|\omega}$.

## IV. Probability of Error

In this paper, we are interested in evaluating the performance of the aforementioned Gibbs sampler, compared to the ML solution. To ease our analysis, we will assume that the ML detector finds the correct transmitted vector. Before we derive the probability of error for the ML detector, we will state a lemma which we will make repeated use of.

**Lemma IV.1** (Gaussian Integral)**.** *Let* $\mathbf{v}$ *and* $\mathbf{x}$ *be independent Gaussian random vectors with distribution* $\mathcal{N}(\mathbf{0}, \mathbf{I}_N)$ *each. Then, if* $1 - 2a^2\eta(1+2\eta) > 0$*,*

$$E\left\{e^{\eta(\|\mathbf{v}+a\mathbf{x}\|^2 - \|\mathbf{v}\|^2)}\right\} = \left(\frac{1}{1-2a^2\eta(1+2\eta)}\right)^{N/2}. \tag{6}$$

*Proof:* See Appendix VIII-A for a detailed proof. ■

Assuming that the vector $\mathbf{s} = -\mathbf{1}$ was transmitted, the ML detector will make an error if there exists a vector $\mathbf{s} \neq -\mathbf{1}$ such that

$$\left\|\mathbf{y} - \sqrt{\frac{\text{SNR}}{N}}\mathbf{H}\mathbf{s}\right\|^2 \leq \left\|\mathbf{y} + \sqrt{\frac{\text{SNR}}{N}}\mathbf{H}\mathbf{1}\right\|^2 = \|\boldsymbol{v}\|^2 .$$

In other words,

$$\begin{aligned} P_e &= \text{Prob}\left(\left\|\mathbf{y} - \sqrt{\frac{\text{SNR}}{N}}\mathbf{H}\mathbf{s}\right\|^2 \leq \|\boldsymbol{v}\|^2\right) \\ &= \text{Prob}\left(\left\|\boldsymbol{v} + \sqrt{\frac{\text{SNR}}{N}}\mathbf{H}(-\mathbf{1}-\mathbf{s})\right\|^2 \leq \|\boldsymbol{v}\|^2\right) , \end{aligned}$$

for some $\mathbf{s} \neq -\mathbf{1}$, which can be formulated as

$$P_e = \text{Prob}\left(\left\|\boldsymbol{v} + 2\sqrt{\frac{\text{SNR}}{N}}\mathbf{H}\boldsymbol{\delta}\right\|^2 \leq \|\boldsymbol{v}\|^2\right) ,$$

for some $\boldsymbol{\delta} \neq 0$. Note that in the above equation $\boldsymbol{\delta}$ is a vector of zeros and $-1$'s. Now using the union bound

$$P_e \leq \sum_{\boldsymbol{\delta}\neq 0} \text{Prob}\left(\left\|\boldsymbol{v} + 2\sqrt{\frac{\text{SNR}}{N}}\mathbf{H}\boldsymbol{\delta}\right\|^2 \leq \|\boldsymbol{v}\|^2\right). \tag{7}$$

We will use the Chernoff bound to bound the quantity inside the summation. Thus,

$$\text{Prob}\left(\left\|\boldsymbol{v}+2\sqrt{\frac{\text{SNR}}{N}}\mathbf{H}\boldsymbol{\delta}\right\|^2 \le \|\boldsymbol{v}\|^2\right) \tag{8a}$$

$$\le E\left\{e^{-\beta\left(\left\|\boldsymbol{v}+2\sqrt{\frac{\text{SNR}}{N}}\mathbf{H}\boldsymbol{\delta}\right\|^2-\|\boldsymbol{v}\|^2\right)}\right\} \tag{8b}$$

$$= \left(\frac{1}{1+8\frac{\text{SNR}\|\boldsymbol{\delta}\|^2}{N}\beta(1-2\beta)}\right)^{N/2}, \tag{8c}$$

where $\beta \ge 0$ is the Chernoff parameter, and where we have used Lemma IV.1 with $\eta = -\beta$ and $a = 2\sqrt{\frac{\text{SNR}\|\boldsymbol{\delta}\|^2}{N}}$, since

$$E\left\{\left(2\sqrt{\frac{\text{SNR}}{N}}\mathbf{H}\boldsymbol{\delta}\right)\left(2\sqrt{\frac{\text{SNR}}{N}}\mathbf{H}\boldsymbol{\delta}\right)^*\right\} = 4\frac{\text{SNR}\|\boldsymbol{\delta}\|^2}{N}\mathbf{I}_N.$$

The optimal value for $\beta$ is $\frac{1}{4}$, which yields the tightest bound

$$\text{Prob}\left(\left\|\boldsymbol{v}+2\sqrt{\frac{\text{SNR}}{N}}\mathbf{H}\boldsymbol{\delta}\right\|^2 \le \|\boldsymbol{v}\|^2\right) \le \left(\frac{1}{1+\frac{\text{SNR}\|\boldsymbol{\delta}\|^2}{N}}\right)^{N/2}. \tag{9}$$

Note that this depends only on $\|\boldsymbol{\delta}\|^2$, the number of nonzero entries in $\boldsymbol{\delta}$. Plugging this into the union bound yields

$$P_e \le \sum_{i=1}^{N}\binom{N}{i}\left(\frac{1}{1+\frac{\text{SNR}i}{N}}\right)^{N/2}. \tag{10}$$

Let us first look at the linear (i.e., $i$ proportional to $N$) terms in the above sum. Thus,

$$\binom{N}{i}\left(\frac{1}{1+\frac{\text{SNR}i}{N}}\right)^{N/2} \approx e^{NH(\frac{i}{N})-\frac{N}{2}\ln\left(1+\frac{\text{SNR}i}{N}\right)},$$

where $H(\cdot)$ is entropy in "nats". Clearly, if $\lim_{N\to\infty}\text{SNR} = \infty$, then the linear terms go to zero (superexponentially fast).

Let us now look at the sublinear terms. In particular, let is look at $i = 1$:

$$N\left(\frac{1}{1+\frac{\text{SNR}}{N}}\right)^{N/2} \approx Ne^{-\text{SNR}/2}.$$

Clearly, to have this term go to zero, we require that $\text{SNR} > 2\ln N$. A similar argument shows that all other sublinear terms also go to zero, and so.[5]

**Lemma IV.2** (SNR scaling)**.** *If SNR* $> 2\ln N$*, then* $P_e \to 0$ *as* $N \to \infty$*.*

[5]Due to space constraints we only present a sketch of this bound. A rigorous proof can be given using the saddle point method, similarly to the proof in the next section.

## V. COMPUTING THE OPTIMAL $\alpha$

Assuming that the vector $\mathbf{s} = -\mathbf{1}$ has been transmitted, the probability of finding this solution *after the Markov chain has mixed* is simply $\pi_{-\mathbf{1}}$, the steady-state probability of being in the all $-1$ state. Clearly, if this probability is exponentially small, it will take exponentially long for the Gibbs sampler to find it. We will therefore insist that the mean of $\pi_{-\mathbf{1}}$ be only polynomially small.

### A. Mean of $\pi_{-\mathbf{1}}$

This calculation has a lot in common with the one given in Section IV. Note that the steady state value of $\pi_{-\mathbf{1}}$ is simply

$$\pi_{-\mathbf{1}} = \frac{e^{-\frac{1}{2\alpha^2}\left\|\mathbf{y}+\sqrt{\frac{\mathrm{SNR}}{N}}\mathbf{H}\mathbf{1}\right\|^2}}{\sum_{\mathbf{s}} e^{-\frac{1}{2\alpha^2}\left\|\mathbf{y}+\sqrt{\frac{\mathrm{SNR}}{N}}\mathbf{H}\mathbf{s}\right\|^2}} \tag{11a}$$

$$= \frac{e^{-\frac{1}{2\alpha^2}\|\boldsymbol{v}\|^2}}{\sum_{\mathbf{s}} e^{-\frac{1}{2\alpha^2}\left\|\boldsymbol{v}+\sqrt{\frac{\mathrm{SNR}}{N}}\mathbf{H}(\mathbf{s}-\mathbf{1})\right\|^2}} \tag{11b}$$

$$= \frac{e^{-\frac{1}{2\alpha^2}\|\boldsymbol{v}\|^2}}{\sum_{\boldsymbol{\delta}} e^{-\frac{1}{2\alpha^2}\left\|\boldsymbol{v}+2\sqrt{\frac{\mathrm{SNR}}{N}}\mathbf{H}\boldsymbol{\delta}\right\|^2}} \tag{11c}$$

$$= \frac{1}{\sum_{\boldsymbol{\delta}} e^{-\frac{1}{2\alpha^2}\left(\left\|\boldsymbol{v}+2\sqrt{\frac{\mathrm{SNR}}{N}}\mathbf{H}\boldsymbol{\delta}\right\|^2-\|\boldsymbol{v}\|^2\right)}}\ , \tag{11d}$$

where $\boldsymbol{\delta}$ is a vector of zeros and ones and the summations (over $\mathbf{s}$ and $\boldsymbol{\delta}$) are over $2^n$ terms.

Now, by Jensen's inequality

$$E\{\pi_{-\mathbf{1}}\} \geq \frac{1}{E\left\{\frac{1}{\pi_{-\mathbf{1}}}\right\}} \tag{12a}$$

$$= \frac{1}{E\left\{\sum_{\boldsymbol{\delta}} e^{-\frac{1}{2\alpha^2}\left(\left\|\boldsymbol{\upsilon}+2\sqrt{\frac{\text{SNR}}{N}}\mathbf{H}\boldsymbol{\delta}\right\|^2-\|\boldsymbol{\upsilon}\|^2\right)}\right\}} \tag{12b}$$

$$= \frac{1}{\sum_{\boldsymbol{\delta}} E\left\{e^{-\frac{1}{2\alpha^2}\left(\left\|\boldsymbol{\upsilon}+2\sqrt{\frac{\text{SNR}}{N}}\mathbf{H}\boldsymbol{\delta}\right\|^2-\|\boldsymbol{\upsilon}\|^2\right)}\right\}} \tag{12c}$$

$$= \frac{1}{1+\sum_{\boldsymbol{\delta}\neq 0}\left(\frac{1}{1+4\frac{\text{SNR}}{N}\|\boldsymbol{\delta}\|^2\frac{1}{\alpha^2}(1-\frac{1}{\alpha^2})}\right)^{N/2}} \tag{12d}$$

$$= \frac{1}{1+\sum_{i=1}^{N}\binom{N}{i}\left(\frac{1}{1+\frac{\beta i}{N}}\right)^{N/2}}\ . \tag{12e}$$

In (12d) we have used Lemma IV.1 and in (12e) we have defined $\beta \triangleq 4\text{SNR}\frac{1}{\alpha^2}(1-\frac{1}{\alpha^2})$. While it is possible to focus on the linear and sublinear terms in the above summation separately, to give conditions for $E\{\pi_{-\mathbf{1}}\}$ to have the form of $1/\text{poly}(N)$, we will be interested in the exact exponent and so will need a more accurate estimate. To do this we shall use saddle point integration. Note that

$$\binom{N}{i}\left(\frac{1}{1+\frac{\beta i}{N}}\right)^{N/2} \approx e^{NH(\frac{i}{N})-\frac{N}{2}\ln\left(1+\frac{\beta i}{N}\right)}\ ,$$

where again $H(\cdot)$ represents the entropy in "nats". And so the summation in the denominator of (12e) can be approximated as a Stieltjes integral:

$$\sum_{i=1}^{N}\binom{N}{i}\left(\frac{1}{1+\frac{\beta i}{N}}\right)^{N/2} \approx N\sum_{i=1}^{N} e^{NH(\frac{i}{N})-\frac{N}{2}\ln\left(1+\frac{\beta i}{N}\right)}\frac{1}{N} \tag{13a}$$

$$\approx N\int_0^1 e^{NH(x)-\frac{N}{2}\ln(1+\beta x)}dx\ . \tag{13b}$$

For large $N$, this is a saddle point integral and can be approximated by the formula

$$\int_0^1 e^{Nf(x)}dx \approx \sqrt{\frac{2\pi}{N|f''(x_0)|}}e^{Nf(x_0)}\ , \tag{14}$$

where $x_0$, is the saddle point of $f(\cdot)$, i.e.,$f'(x_0)=0$. In our case,

$$f(x) = -x\ln x - (1-x)\ln(1-x) - \frac{1}{2}\ln(1+\beta x)\ ,$$

and so

$$f'(x) = \ln\frac{1-x}{x} - \frac{1}{2}\frac{\beta}{1+\beta x}\ .$$

In general, it is not possible to solve for $f'(x_0)=0$ in closed form. However, in our case, if we assume that $\beta = 4\text{SNR}\frac{1}{\alpha^2}(1-\frac{1}{\alpha^2}) \gg 1$ (which is true since the SNR grows at least logarithmically), then it is not too hard to verify that the saddle point is given by

$$x_0 = e^{-\frac{\beta}{2}}\ . \tag{15}$$

And hence $f(x_0) =$

$$\begin{aligned} & -e^{-\frac{\beta}{2}}\ln e^{-\frac{\beta}{2}} - (1-e^{-\frac{\beta}{2}})\ln(1-e^{-\frac{\beta}{2}}) - \frac{1}{2}\ln(1+\beta e^{-\frac{\beta}{2}}) \\ & \approx \frac{\beta}{2}e^{-\frac{\beta}{2}} + e^{-\frac{\beta}{2}} - \frac{1}{2}\beta e^{-\frac{\beta}{2}} = e^{-\frac{\beta}{2}}\ , \end{aligned}$$

and further plugging $x_0$ into $f''(x) = -\frac{1}{x} - \frac{1}{1-x} - \frac{1}{2}\frac{\beta^2}{(1+\beta x)^2}$, yields

$$f''(x_0) \approx -e^{\frac{\beta}{2}} - 1 + \frac{1}{2}\beta^2 \approx -e^{\frac{\beta}{2}}\ . \tag{16}$$

Replacing these into the saddle point expression in (14) show that

$$\sum_{i=1}^{N}\begin{pmatrix} N \\ i \end{pmatrix}\left(\frac{1}{1+\frac{\beta i}{N}}\right)^{N/2} \approx \sqrt{2\pi/N}\exp\left(Ne^{-\frac{\beta}{2}} - \frac{\beta}{4}\right)\ . \tag{17}$$

We want $E\left\{\pi_{-\mathbf{1}}\right\}$ to behave as $\frac{1}{N^\zeta}$ and according to (12) this means that we want the expression in (17) to behave as $N^\zeta$. Let us take

$$e^{Ne^{-\frac{\beta}{2}}} = N^\zeta\ .$$

Solving for $\beta$ yields

$$\beta = 4\text{SNR}\frac{1}{\alpha^2}\left(1-\frac{1}{\alpha^2}\right) = 2\left(\ln N - \ln\ln N - \ln\zeta\right). \tag{18}$$

Incidentally, this choice of $\beta$ yields $e^{-\frac{\beta}{4}} \approx \frac{1}{\sqrt{N}}$, and so we have the following result.

**Lemma V.1** (Mean of $\pi_{-\mathbf{1}}$)**.** *If $\alpha$ is chosen such that*

$$\frac{\alpha^2}{1-\frac{1}{\alpha^2}} = \frac{2\mathit{SNR}}{\ln N - \ln\ln N - \ln\zeta}\ , \tag{19}$$

*then*

$$E\left\{\pi_{-\mathbf{1}}\right\} \geq N^{-\zeta}\ . \tag{20}$$

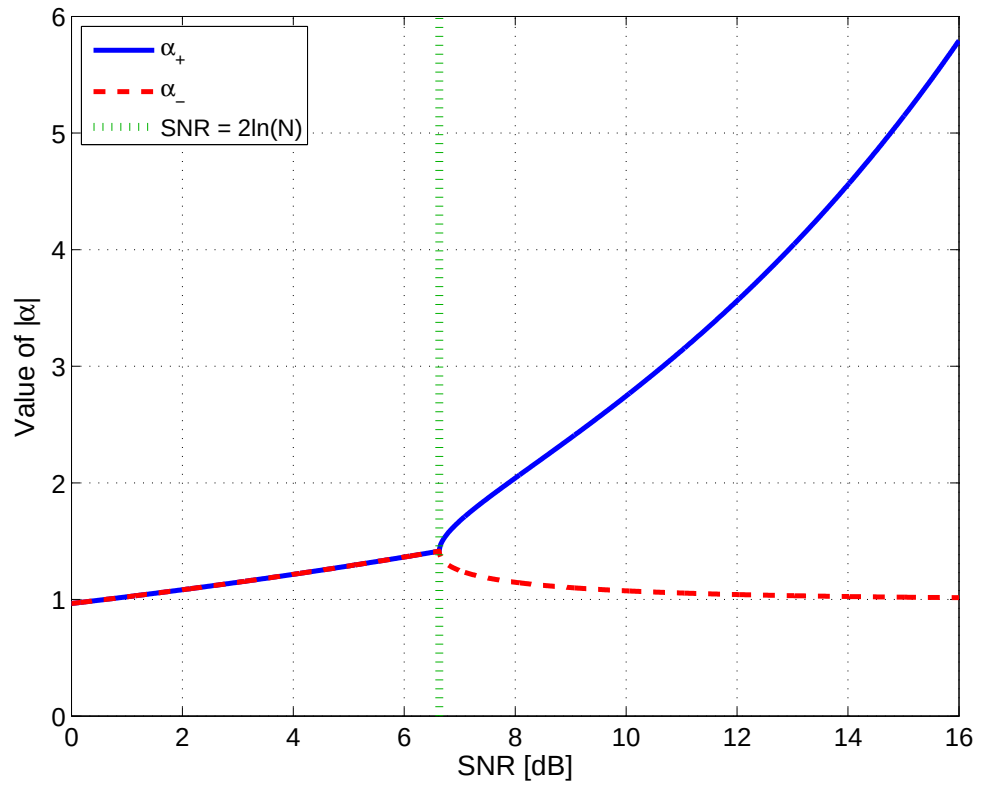

Figure 1: Value of $\alpha$ vs. SNR for system size $N = 10$.

## B. *Value of* $\alpha$

Note that from (12e) it is clear that the larger $\beta$ is, the larger $\pi_{-\mathbf{1}}$ is. Therefore, the range of $\alpha$ that gives a polynomially small probability to $\pi_{-\mathbf{1}}$ is

$$\frac{\alpha^2}{1-\frac{1}{\alpha^2}} \geq \frac{2\mathrm{SNR}}{\ln N - \ln\ln N - \ln\zeta}. \tag{21}$$

It can be shown that in the regime, $\mathrm{SNR} > 2\ln N$, the above quadratic inequality in $\alpha$ has two positive real solutions, $\alpha_+ \geq \alpha_-$, and that the inequality holds for all $\alpha \in [\alpha_-, \alpha_+]$.

We know that, the larger $\alpha$ is, the faster the Markov chain mixes.[6] Therefore it is reasonable that we choose the largest permissible value for $\alpha$, i.e., $\alpha_+$.

Figures 1 and 2 show the values of $\alpha_+$ and $\alpha_-$ as a function of SNR for systems with $N = 10$ and $N = 50$, when we have $\zeta = 1/\ln N$.

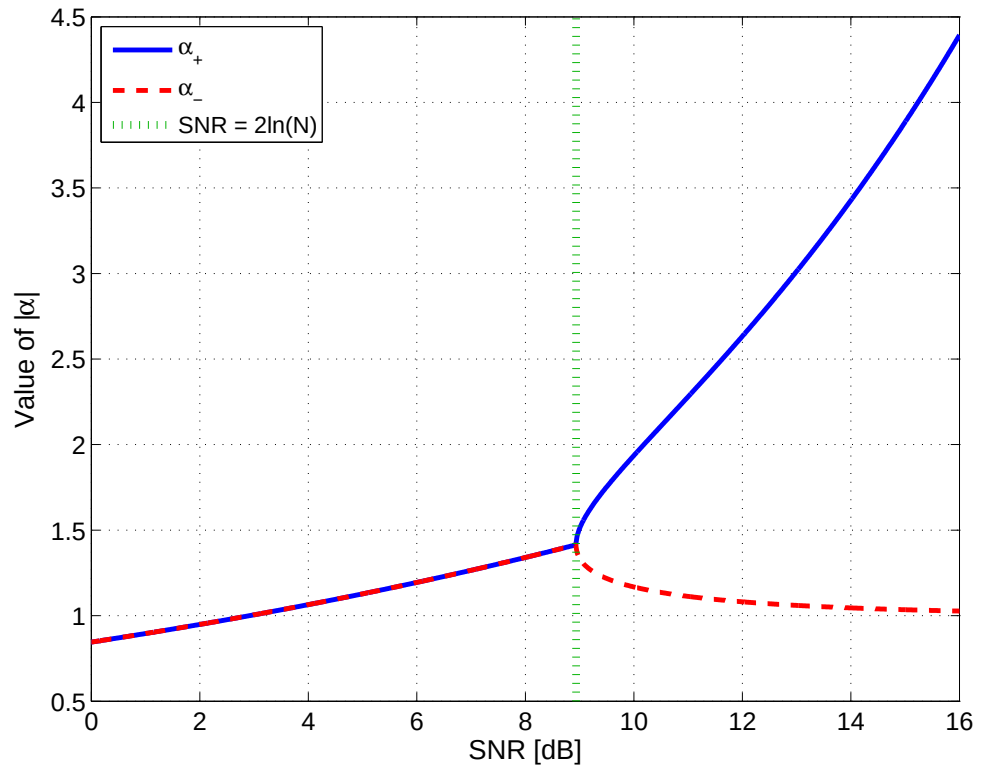

Figure 2: Value of $\alpha$ vs. SNR for system size $N = 50$.

[6]In general, there is a trade-off between faster mixing time of the Markov chain (due to an increase of $\alpha$) versus slower encountering the optimal solution in steady-state. In fact, at infinite temperature our algorithm reduces to a random walk in a hypercube which mixes in $O(N \ln N)$ time.

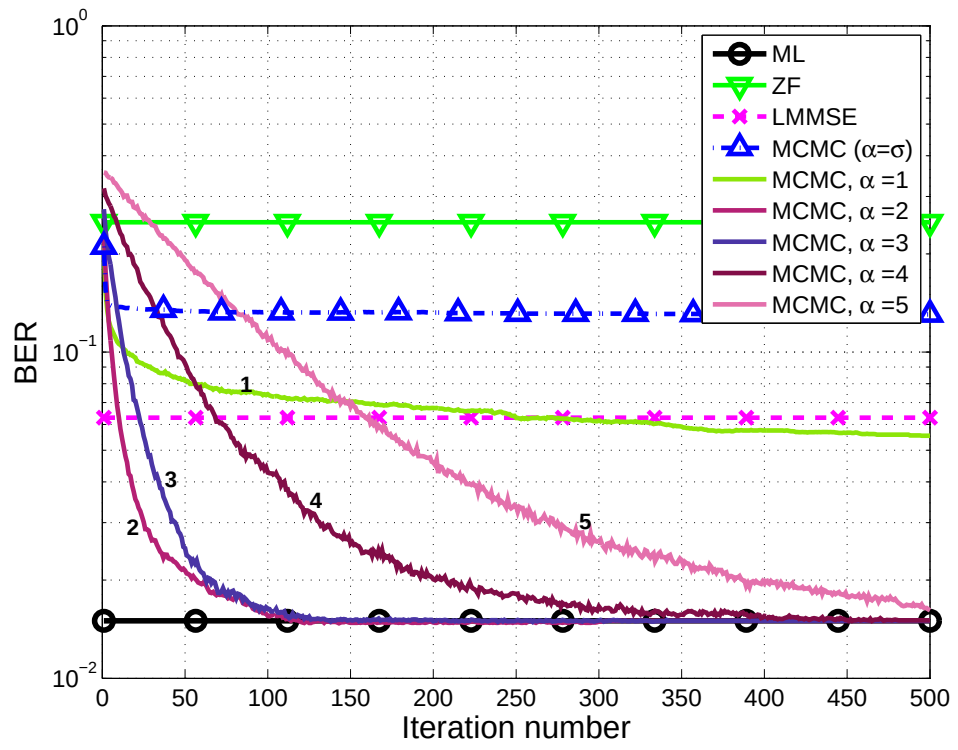

Figure 3: BER vs. iterations, $10 \times 10$. SNR = 10 dB.

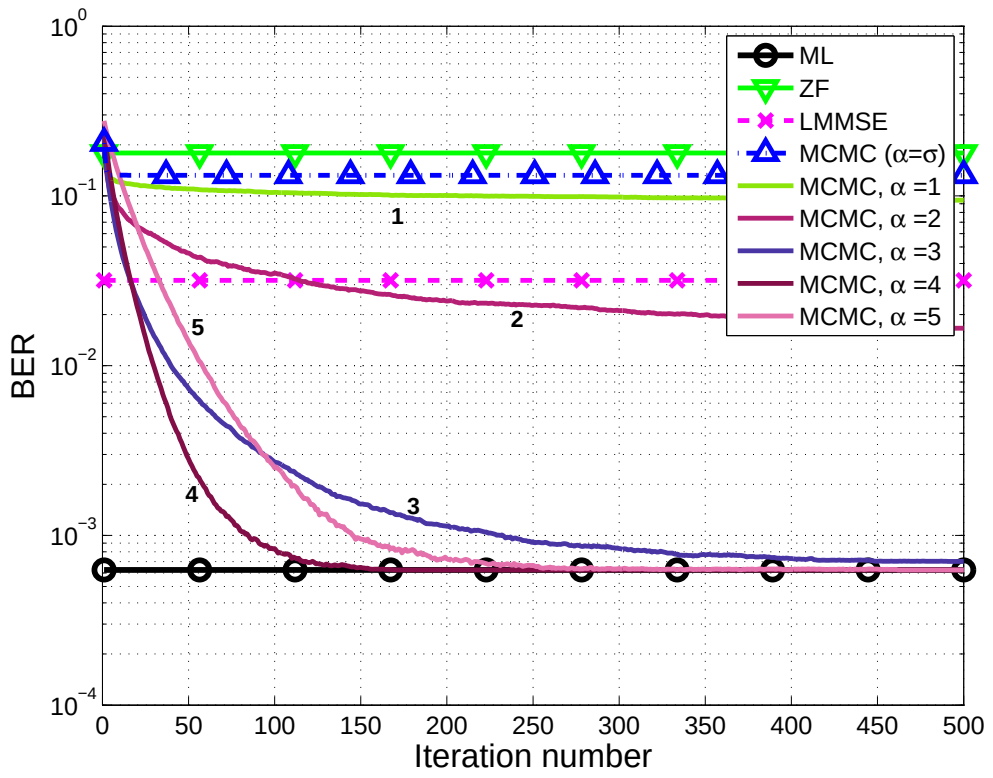

Figure 4: BER vs. iterations, $10 \times 10$ system. SNR = 14 dB.

### C. *Mixing time of Markov Chain*

One open question is whether the Markov chain is rapidly mixing when using the strategy above for choosing $\alpha$. This is something we are currently investigating, and the simulations presented in Section VI seem to indicate that this is the case. Furthermore, the simulations also suggest that the computed value of $\alpha$ is very close to the optimal choice, even in the case where the condition SNR $> 2\ln(N)$ is not satisfied.

## VI. Simulation Results

In this section we present simulation results for a MIMO $N \times N$ system with a full square channel matrix containing i.i.d. Gaussian entries. In Fig. 3 and Fig. 4 the Bit Error Rate (BER) of the Gibbs sampler, initialized with a random $\mathbf{s}$, has been evaluated as a function of the number of iterations in a $10 \times 10$ system using a variety of $\alpha$ values. Thereby, we can inspect how the parameter $\alpha$ affects the convergence rate of the Gibbs sampler. The performance of the Maximum Likelihood (ML), the Zero-Forcing (ZF), and the Linear Minimum Mean Square Error (LMMSE) detector has also been plotted, to ease the comparison of the Gibbs sampler with these. It is seen that the Gibbs sampler outperforms both the ZF and the LMMSE detector after only a few iterations in all the presented simulations, when the tuning parameter $\alpha$ is chosen properly. Furthermore, it is observed that the parameter $\alpha$ has a huge influence on the convergence rate and that the Gibbs sampler converges toward the ML solution as a function of the iterations.[7] The optimal value of $\alpha$ (in terms of convergence rate) is quite close to the theoretical values from Fig. 1 of $\alpha_+ = 2.7$ and $\alpha_+ = 4.6$ at SNR's at $10$ and $14dB$, respectively. It is also observed that the performance of the Gibbs sampler is significantly deteriorated if the temperature parameter is chosen based on the SNR (and thereby on the noise variance), such that $\alpha = \sigma \triangleq 1/\text{SNR}$. Thus, the latter strategy is clearly not a wise choice.

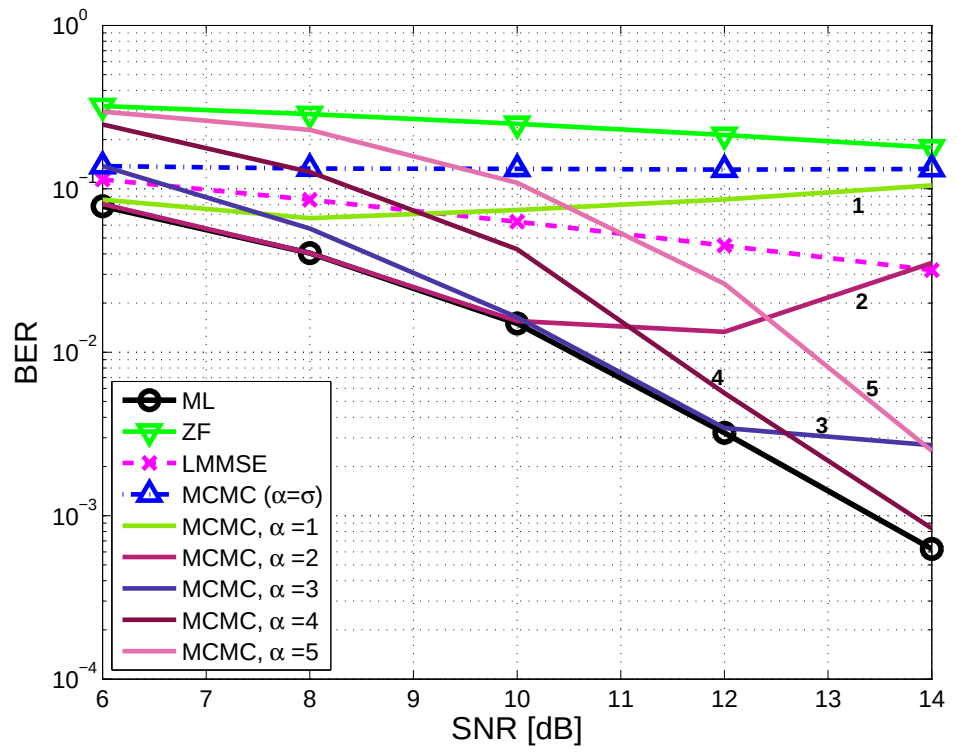

Figure 5: BER vs. SNR, $10 \times 10$. Number of iterations, $k = 100$.

Figure 5 shows the BER performance for the MCMC detector for fixed number of iterations, $k = 100$. From the figure we see that the SNR has a significant influence on the optimal choice of $\alpha$ given a fixed number of iterations.

The performance of the Gibbs sampler is also shown for a $50 \times 50$ system, which represents a ML decoding problem of huge complexity where an exhaustive search would require $2^{50} \approx 10^{15}$ evaluations. For this problem even the sphere decoder has an enormous complexity under moderate SNR.[8] Therefore, it has not been possible to simulate the performance of this decoder within a reasonable time and we have therefore "cheated" a little by initializing the radius of the sphere to the minimum of either the norm of the transmitted symbol vector or the solution found by the Gibbs sampler. This has been done in order to evaluate the BER performance of the optimal detector. Figure 6 shows the BER curve as a function of the iteration number, while Figure 7 illustrates the BER curve vs. the SNR. From Figure 6 we see that there is a quite good correspondence between the simulated $\alpha$ and the theoretical value $\alpha_+ = 2.6$ obtained from Figure 2. The average complexity (MAC pr. symbol vector) of the Gibbs sampler having a BER performance comparable with the ML detector is shown in Table I. The SD has been included as a reference.[9] It is observed that the complexity of the Gibbs sampler is not affected by the SNR as much as the SD.

[7] It should be noted that the way we decode the symbol vector to a given iteration, is to select the symbol vector which has the lowest cost function in all the iterations up to that point in time.

[8] In fact, it can be shown that, for SNR $= \mathcal{O}(\ln N)$, the lower bound on the complexity of the sphere decoder obtained in [6] is exponential.

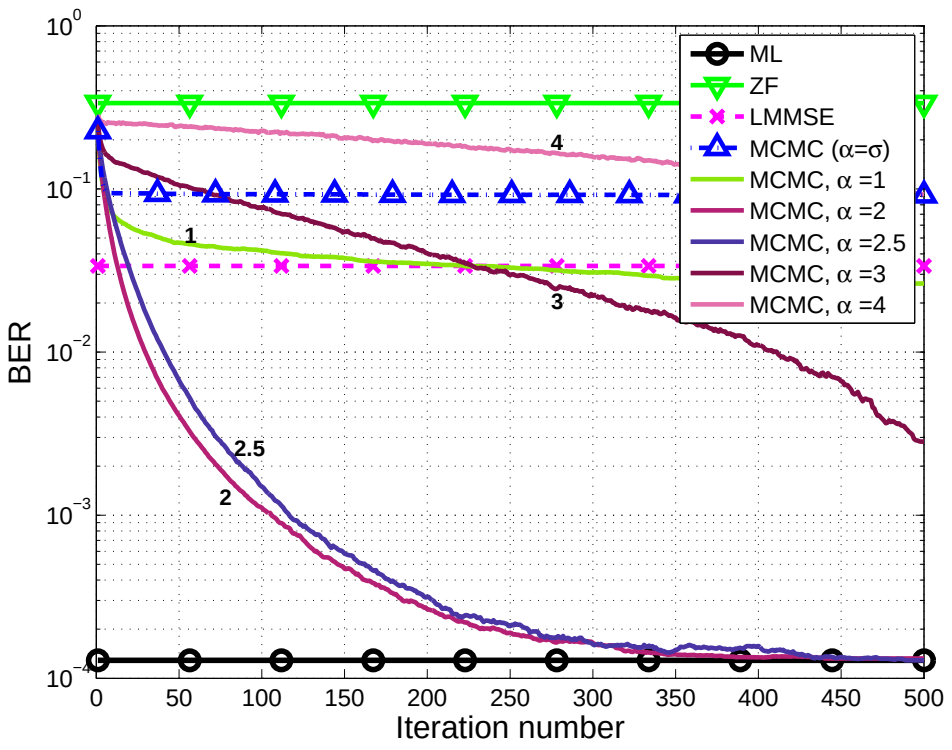

Figure 6: BER vs. iterations, $50 \times 50$ system. SNR = 12 dB.

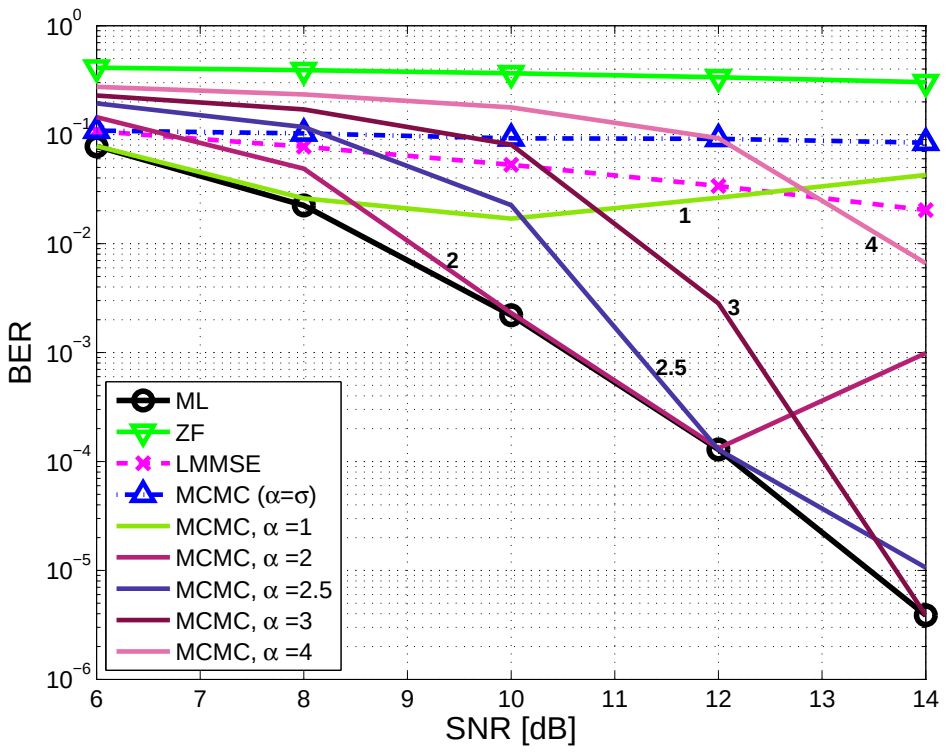

Figure 7: BER vs. SNR, $50 \times 50$ system. Num. of iter., $k = 500$.

Table I: Complexity of SD and Gibbs Sampler (GS).

| $N$ | Method | SNR 6 dB | 10 dB | 14 dB |
|---|---|---|---|---|
| 10 | GS | $9.8 \cdot 10^3$ | $10.9 \cdot 10^3$ | $16.4 \cdot 10^3$ |
| | SD | $10.0 \cdot 10^3$ | $1.7 \cdot 10^3$ | $1.5 \cdot 10^3$ |
| 50 | GS | $7.6 \cdot 10^5$ | $9.5 \cdot 10^5$ | $10.6 \cdot 10^5$ |
| | SD | $\gg 1.9 \cdot 10^9$ | $\gg 1.9 \cdot 10^9$ | $37.7 \cdot 10^5$ |

## VII. Conclusion

In this paper we considered solving the integer least-squares problem using Monte Carlo Markov Chain Gibbs sampling. The novelty of the proposed MCMC method is that, unlike simulated annealing techniques, we have a fixed temperature parameter in all the iterations, with the property that after the Markov chain has mixed, the probability of encountering the optimal solution is only polynomial small (i.e. not exponentially small). We further compute the optimal (here largest) value of the temperature parameter that guarantees this. Simulation results indicate the sensitivity of the method to the choice of the temperature parameter and show that our computed value gives a very good approximation to its optimal value. Investigating whether the Markov chain mixes in polynomial time for this choice of temperature parameter is currently under investigation.

[9] It has not been possible to simulate the SD for a $50 \times 50$ system when $SNR \leq 10dB$ and, therefore, the complexity of $SNR = 12dB$ has been used a lower bound.

## VIII. Appendix

### A. Proving Lemma IV.1

**Lemma IV.1** (Gaussian Integral) *Let* $\mathbf{v}$ *and* $\mathbf{x}$ *be independent Gaussian random vectors with distribution* $\mathcal{N}(\mathbf{0}, \mathbf{I}_N)$ *each. Then*

$$E\left\{e^{\eta(\|\mathbf{v}+a\mathbf{x}\|^2-\|\mathbf{v}\|^2)}\right\} = \left(\frac{1}{1-2a^2\eta(1+2\eta)}\right)^{N/2} . \tag{22}$$

**Proof:** In order to determine the expected value we compute the multivariate integral

$$\begin{aligned} & E\left\{e^{\eta(\|\mathbf{v}+a\mathbf{x}\|^2-\|\mathbf{v}\|^2)}\right\} \\ &= \int \frac{d\mathbf{x}d\mathbf{v}}{(2\pi)^N} e^{-\frac{1}{2}\left[\,\mathbf{v}^T,\ \mathbf{x}^T\,\right]\begin{bmatrix} \mathbf{I}_N & -2a\eta\mathbf{I}_N \\ -2a\eta\mathbf{I}_N & (1-2a^2\eta)\mathbf{I}_N \end{bmatrix}\begin{bmatrix}\mathbf{v}\\ \mathbf{x}\end{bmatrix}} \\ &= \frac{1}{\det^{N/2}\begin{bmatrix} 1 & -2a\eta \\ -2a\eta & 1-2a^2\eta \end{bmatrix}} = \left(\frac{1}{1-2a^2\eta(1+2\eta)}\right)^{N/2} . \end{aligned}$$

Thus, Lemma IV.1 has hereby been proved. ■